%% file: main.tex
\documentclass[10pt,conference]{IEEEtran}

\usepackage{amsmath,amssymb,amsfonts,bm}
\usepackage{algorithm}
\usepackage{algorithmic}
\usepackage{graphicx}
\usepackage{textcomp}
\usepackage{xcolor}
\usepackage{siunitx}
\usepackage[capitalize]{cleveref}
\usepackage{booktabs}
\usepackage{tabularx}
\usepackage{pgfplots}
\usepackage{comment}

\DeclareSIUnit\hours{hours}
\usepackage{pifont}
\newcommand{\cmark}{\ding{51}} 
\newcommand{\xmark}{\ding{55}} 

\usepackage[style=ieee, maxcitenames=2, mincitenames=1, citestyle=numeric-comp,backend=bibtex]{biblatex}
\AtEveryBibitem{
 	\clearfield{url}
 	\clearfield{doi}
\ifentrytype{inproceedings}{
 	\clearlist{address}
 	\clearlist{publisher}
 	\clearname{editor}
 	\clearlist{organization}
 	\clearfield{pages}
 	\clearlist{location}
	\clearfield{volume}
	\clearfield{series}
	\clearfield{eprint}
	\clearfield{eprinttype}
 }{}
 }
\def\BibTeX{{\rm B\kern-.05em{\sc i\kern-.025em b}\kern-.08em
    T\kern-.1667em\lower.7ex\hbox{E}\kern-.125emX}}

\usepgfplotslibrary{groupplots}
\usetikzlibrary{positioning}
\pgfplotsset{compat=newest}
\pgfplotsset{every axis legend/.append style={%
cells={anchor=west}}
}
\pgfplotsset{every axis/.append style={
                    label style={font=\tiny},
					tick label style={font=\tiny},
					legend style={font=\tiny}
                    }}

\IEEEoverridecommandlockouts
\begin{document}

\title{Mobile Network Control with a World Model\\
}

\author{
 Maxime Bouton$^*$, Ioanna Mitsioni$^*$, Simon Lindståhl, Jaeseong Jeong
\\ Ericsson Research, Stockholm, Sweden
\\ \small{ $^*$ Denotes equal contribution}
\thanks{This work has been accepted for publication in the 2026 IEEE/IFIP Network Operations and Management Symposium (NOMS). The final published version will be available on IEEE Xplore.}
}

\maketitle

\begin{abstract}
The increasing complexity of mobile networks necessitates intelligent and dynamic control strategies for efficient, energy-conserving management. We propose a world model-based approach for network control that enables adaptive configuration of crucial parameters. 
The world model is trained from historical data and predicts the impact of its actions on future network states. Our controller leverages the model's uncertainty estimate to robustly find optimal network configuration changes. Furthermore, the optimization objective can be changed dynamically without model retraining.
We demonstrate the effectiveness of the approach in simulated closed-loop control of a mobile network energy-saving feature. Our results show improved performance in balancing energy savings with quality of service, compared to traditional methods and reinforcement learning approaches. Finally, we show the world model performance on real network data from, and evaluate counterfactual actions proposed by the controller under various throughput constraints.

\end{abstract}

\section{Introduction}
\label{sec:intro}
\input{sections/intro}

\section{Related work}
\label{sec:related}
\input{sections/related}

\section{Cell sleep mode control}
\label{sec:system}
\input{sections/system}

\section{World model-based control}
\label{sec:approach}
\input{sections/approach}

\section{Simulation Experiments}
\label{sec:simulation}
\input{sections/simulation}

\section{Live Network Data Experiments}
\label{sec:data}
\input{sections/real_data}


\section{Conclusion}
\label{sec:conclusion}

\input{sections/conclusion}

\printbibliography

\newpage 

\label{sec:appendix}
\input{sections/appendix}

\end{document}

%% file: sections/intro.tex
Rapid growth in the usage, density, and complexity of mobile networks poses significant challenges for operators in terms of service assurance and energy efficiency. With rising energy costs and sustainability goals, optimizing network configuration is crucial.
Intelligent automated network configuration solutions can enhance quality of service, save energy, and reduce manual intervention and operational costs.
However, optimal control of network configuration parameters at large scale with thousands of Base Stations (BSs) is challenging due to hard-to-model, dynamic and noisy factors, such as network usage and radio propagation conditions.

Previous works address this problem through hand-engineered strategies, optimization algorithms, and reinforcement learning (RL)~\cite{salahdine}. 
Hand-engineered strategies consist of rule-based methods relying on expert knowledge, and may lack adaptability to site-specific information and temporal patterns. 
Optimization methods, with mathematical models of the network and user behavior, often resort to simplifying assumptions to reduce computational costs. 
On the other hand, RL has shown promise in certain management problems, such as antenna tilt configuration~\cite{mendo2023multi} and energy efficiency optimization~\cite{ye2019drag, gsma, pujol2021deep}. However, RL requires extensive training using network simulators or offline data and often needs to be retrained when changing objectives. In a mobile network, different parts of the network may require different behaviors, which can substantially increase the engineering effort required for solutions that learn specific policies.


\begin{figure}[!t]
    \centering
    \includegraphics[width=\linewidth]{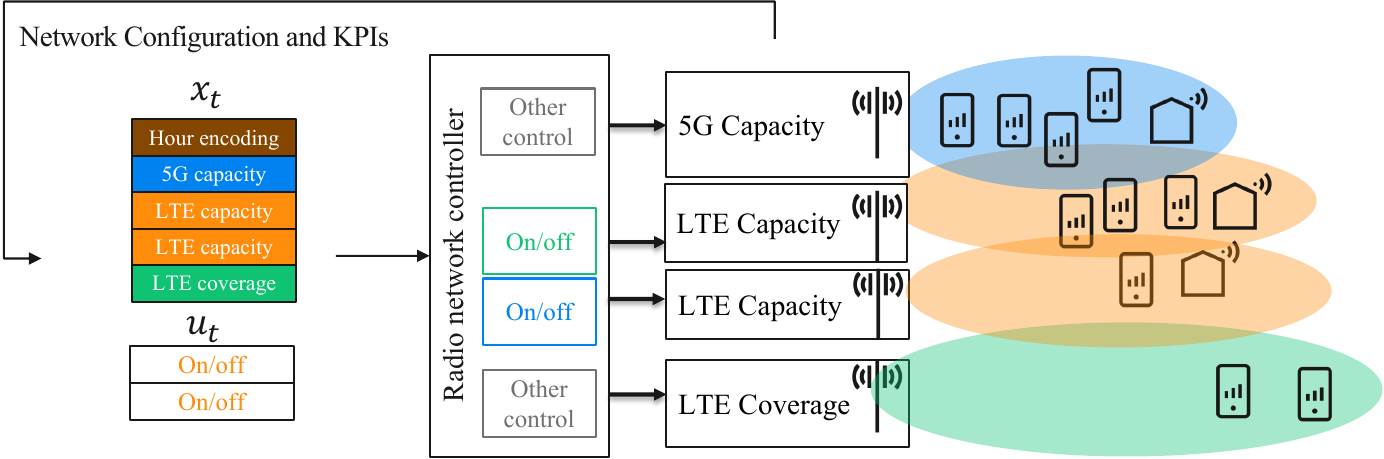}
    \caption{Illustration of our capacity cell sleep control problem. Mobile networks are composed of different frequency band covering the same area. Here, we control the on/off state of one or more LTE frequency bands every hour.}
    \label{fig:network}
\end{figure}

We propose a model-based controller where a world model (WM) learns from available 4G/5G mobile network data. Data-driven WMs have been extensively studied in model-based RL to increase sample efficiency and planning horizon in robotics and game environments~\cite{chua2018deep, hafner2023mastering}.
However, direct application to network data can be challenging since their accuracy relies heavily on datasets with good coverage of the feature and action spaces. In a live network, exploratory and diverse actions may be unsafe, resulting in datasets with poor coverage, which can in turn lead to models insensitive to actions. This is exacerbated by the tabular nature of network data, which often lack obvious structure and consist of continuous, binary, and semi-continuous features. As a consequence, existing WM formulations do not directly apply to Radio Access Network (RAN) control, since they cannot capture the complex interactions of the features and effectively model the effect of different actions. 
Additionally, network data can be noisy and partially observable. For some network metrics, such as user throughput, external factors and unobserved configurations can influence the measurements, making them difficult to model. To mitigate these effects, we propose an architecture with improved feature modeling that can model uncertainty and study how it impacts the resulting control strategy.

This work makes three key contributions. First, we propose a novel feature modeling architecture that improves the action sensitivity of WMs or tabular, heterogeneous, and partially observable mobile network data. Second, we formulate RAN optimization as a model-based control problem, decoupling model learning from the policy to enhance its flexibility. Lastly, we incorporate uncertainty modeling to handle noisy and partial observability and analyze its effect on control performance.
We demonstrate our approach on the use case of controlling the sleep state of capacity cells in a 5G network that requires balancing hardware deactivation for energy savings with maintaining sufficient quality of service. 


We first analyze the performance of the proposed model architecture in terms of quality of predictions and show that it handles the complex nature of network data. Then, we evaluate the closed-loop performance of a WM-based controller in a high fidelity simulation environment. We show that the proposed method outperforms RL alternatives trained in the same simulation environment and a hand-engineered baseline. 
Finally, we analyze the performance of our WM on data from a live 5G network. We measure the prediction performance of the model and evaluate the impact of actions proposed by the controller, thus giving us a coarse estimate of potential energy savings.
These results establish a first step towards introducing the WM-based control paradigm in network automation, improving how modern networks are managed for optimal service quality and energy efficiency.

%% file: sections/related.tex




In network management, dynamic control of RAN features for energy optimization is well-studied, as covered in the 5G RAN survey~\cite{lopez2022survey} and in~\cite{salahdine} for ultra-dense networks in particular. Typical heuristics-based approaches include pre-defining sleep schedules or simple rule-based approaches. Cell sleep modes~\cite{salem} are of particular importance due to their large measured energy savings.

In recent years, there has been a shift in paradigm towards data-driven methods for RAN control, with the most prominent one being reinforcement learning (RL). 
For example, in the problem of antenna tuning using RL, where an RL agent proposes action changes every day~\cite{mendo2023multi} and downlink power control as in \cite{saeidian2020downlink}.
In~\cite{Wang2024}, RL is used to improve energy efficiency by comparing different state representations, in ~\cite{Gui2025} multi-agent RL 
controls both BSs and edge servers. The authors in \cite{el2022energy} utilize RL in heterogeneous networks (HetNets) for multi-level sleep, while in~\cite{Fourati2024} RL is applied for energy savings in the Heterogeneous Cloud 5G RAN. Similar to this work, the authors in~\cite{Bordin2025DRL} train an agent to activate/deactivate Radio Frequency frontends. In contrast to our work, these techniques require re-training when the objective needs to be tweaked. On the model-based side, \cite{Pan2025} use model-based RL in a joint sleep mode and power control method to improve sample efficiency and convergence compare to model-free RL. 

Another popular approach is predicting future network states and combining them with control methods. In~\cite{jiang2018data}, network cell zooming is modeled through a grey-box framework and approximately solved as a knapsack problem. The work in~\cite{Masero2020}, introduces a coalitional Model Predictive Control (MPC) approach with a linear model for energy-efficient operations in HetNets. The authors in~\cite{Balasingam2024} apply a similar logic ours with a data-driven constrained MPC approach, however the task is application-level service assurance with slicing. On the other hand, the authors in~\cite{vallero2019greener} and~\cite{s25164978} use similar data-driven models to ours to predict user needs for the purpose of energy optimization, treating cell sleep and resource allocation respectively. However, they employ their models only for simple heuristic policies and do not take the effect of proposed actions into account for predictions. To the best of our knowledge, ours is the first work that relies on data-driven model-based optimization for RAN control.

Data-driven dynamics models are a mature research field with applications in system control and finances,
but the concept of world models to enable reasoning in AI was codified in the seminal paper~\cite{ha2018world} and has proven useful in planning for diverse domains \cite{hafner2023mastering, hansen2023td}. An important extension are models which can express their own confidence or uncertainty about its knowledge by predicting not only a point estimate but a distribution of possible outcomes \cite{chua2018deep}.
Compared to these approaches, our work is novel in its modeling of RAN-specific dynamics and constraints.


%% file: sections/system.tex
Mobile networks comprise multiple BSs, each equipped with antennas operating on various carrier frequencies for the same sectors, as seen in \cref{fig:network}. This redundancy ensures sufficient bandwidth and quality of service during high-demand periods.
During low-demand periods, disabling redundant frequencies can significantly reduce energy consumption. This method is commonly referred to as cell sleep, where a cell refers to a specific carrier frequency in a specific sector, and is a core part of energy saving techniques in mobile networks. However, disabling a frequency leads to users moving to active ones, potentially congesting the cell.

In 4G/5G service management and orchestration (SMO), cell sleep can be controlled using a cell sleep mode (CSM) feature, with cells divided into always-on \emph{coverage} or sleep candidate \emph{capacity} cells. CSM operates by comparing physical resource block (PRB) utilization of a cell to pre-defined “sleep” and “wake-up” thresholds~\cite{3gpp-tr-36.927}. 
Whenever the utilization of the considered capacity cell drops below the sleep threshold, the cell initiates sleep procedures and is unutilized. During sleep, the utilization of the capacity cell remains at zero and wake-up is instead triggered when the utilization of its associated coverage cell exceeds the wake-up threshold. If more fine-grained control is desired, an rApp can reconfigure thresholds to match current traffic patterns. The sleep and wake-up rules are typically implemented in the BS  and operate on timescales of seconds or faster, but the rApp can only reconfigure thresholds once every quarter-hour or slower. 

The CSM feature benefits from centralized network control schemes that tune thresholds for each sector and adapt them to slow variations in traffic demand, while letting rule-based software in the BS wake-up the cells quickly if needed.
Setting those thresholds using rule-based methods is practical, but for more advanced control schemes, the continuous nature of the cell sleep thresholds makes for unwieldy and counter-intuitive control actions. 
Instead, we operate using \emph{on/off} rules, where the capacity cell is controlled to be either fully awake (sleep and wake-up thresholds are set to zero) or “near-fully” asleep (sleep and wake-up thresholds are set close to their maximal value).
By doing so, we can exert a high degree of control during coarse time granularity (hourly or quarterly) while allowing for quick responses to unexpected scenarios at finer time granularity enabled by software running in the BSs.

%% file: sections/approach.tex
In this section, we present our WM-based approach for network control, describe the WM design and training, its uncertainty estimation, the use of multi-task modeling for tabular data, and its integration with planning algorithms to form the overall control scheme. More details about the WM can also be found in the Appendix.

\subsection{Data-driven world models}
A world model (WM) represents system dynamics by mapping a current state to a next state and predicting outcomes under configuration changes. While it can be physics-based, this is not always feasible in telecommunication problems where the scale of the network and the uncertainties in the environment (e.g., user mobility, radio propagation conditions) make it difficult to have a closed-form model that generalizes to different conditions. Instead, we can learn a data-driven WM from system interactions, either in simulation or online. 

Formally, given a state measurement $x_t\in\mathbb{R}^n$ and a configuration change (control input) $u_t\in \mathbb{R}^k$ of the system at time $t$, a deterministic WM calculates the next state  $x_{t+1}\in \mathbb{R}^n$ based on a function $h(x)$: $x_{t+1}=h(x_t,u_t)$.
When the model is learned from data, we approximate the function $h(x)$ with a parametric model $f_\theta(x)$ (e.g., a neural network) where $\theta$ represents the parameters of the model:
\begin{equation}
    x_{t+1}\sim f_\theta(x_t,u_t).
\label{eq:det_wm}
\end{equation}

In mobile network control, $x_t$ is typically a vector of key performance indicators (KPIs) and $u_t$ are controllable network configurations. In this paper, we model $u_t \in \{0,1\}^k$ as on/off control of capacity cells, where $k$ is the number of capacity cells in a given sector. This problem formulation is easily extendable to different use cases and network configuration parameters, such as transmitted power or antenna tilt angles. 

Stochastic WMs follow the same principles, but learn probabilistic state transitions. They capture environmental uncertainty and account for noise, partial observability, and multiple plausible futures, making them particularly useful in complex, real-world domains. Mathematically, a stochastic WM can be described as learning a conditional probability distribution $p(x_{t+1} \mid x_t, u_t)$ that is used to predict future states:
\begin{equation}
    x_{t+1} \sim f_\theta(\ \cdot \mid x_t, u_t).
\label{eq:sto_wm}
\end{equation}

A trained WM can make predictions about the system's evolution based on the desired input using Equations \eqref{eq:det_wm} or \eqref{eq:sto_wm}. To predicting multiple  steps ahead, the model can be queried auto-regressively by using the predicted next states as inputs. Without loss of generality, we assume that the state $x$ comprises sequences of historical data and that the WM is a sequence-to-sequence model that performs multi-step ahead predictions auto-regressively until horizon $H$. A loss function is then applied to penalize the model for the prediction from $t+1$ to $t=H$ (Mean Squared Error for deterministic models and Negative Log Likelihood for stochastic WMs). In our experiments we use different values of $H$ at training time and test time as detailed in \cref{sec:simulation,sec:data}.

\medskip

\noindent{\textbf{Uncertainty estimation and propagation:}}
In data-driven modeling, uncertainty is usually decomposed into two parts: the inherent stochasticity of the system (aleatoric uncertainty) and the uncertainty of the approximation due to insufficient data (epistemic uncertainty). Stochastic models capture aleatoric uncertainty, while epistemic uncertainty can be estimated by using ensembles~\cite{chua2018deep}. In this work, we construct ensembles the same way as \citeauthor{chua2018deep}, resulting in $M$ neural networks that are trained independently to model the system dynamics.  
The ensemble's predictive distribution is then:
\begin{equation}
    p(x_{t+1} \mid x_t, u_t) \approx \frac{1}{M} \sum_{m=1}^M f_{\theta_m}(x_t, u_t).
\end{equation}

Deterministic ensembles capture epistemic uncertainty through the disagreement across their members, while stochastic ensembles additionally reflect aleatoric uncertainty by sampling within each model’s distribution. We calculate both uncertainty types as variances \cite{bulte2025axiomatic}. At inference time, uncertainty is propagated, 
thus preventing overconfident predictions and enabling more robust long-term control under uncertainty. 

\subsection{Multi-task model}
In most previous works involving WMs, state features $x$ are continuous, making the task a regression problem~\cite{chua2018deep}. However, in domains with switching dynamics, such as networks with CSM, some features are binary (sleep state) and force others to be semi-continuous. For example, the PRB utilization and throughput of a cell that is changing to a sleeping state will suddenly become discontinuous. Although a rich dataset might allow the WM to capture this switching behavior, it is a difficult and data-inefficient solution as switching samples are more infrequent than steady-state ones.

To overcome this problem, we propose an approach for learning WMs with an architecture that handles continuous, binary, and semi-continuous features explicitly, making the problem a multi-task one. For the multi-task formulation, we model the state $x_t$ as a combination of its different features: 
$$
x_t = \begin{bmatrix}
    x_t^c &
    x_t^b &
    x_t^{sc}
\end{bmatrix} ^T
$$
where $x_t^c \in \mathbb{R}^{n_c}$ are the continuous features, $x_t^b \in \{0,1\}^{n_b}$ are the binary features and $x_t^{sc} \in \mathbb{R}^{n_{sc}}_+$ are the semi-continuous features.
The multi-task states and control inputs are embedded separately and processed by a sequence-to-sequence backbone before being split to task-specific output heads. 
We then encourage the model to learn the dependency between binary and semi-continuous features by applying an activation mask on the semi-continuous prediction based on the binary prediction. This approach ensures consistency between the two sets of variables and makes their dependency explicit while learning.

The multi-task model is still trained end-to-end like its regression counterparts but its optimization criterion encompasses the different feature types and forms a weighted multi-task loss function:
\begin{equation}
\mathcal{L} = \sum_{t=1}^{T+H} [\lambda_c\mathcal{L}_\text{c}(x_t^c, \hat{x}_{t}^c) + \lambda_b \mathcal{L}_\text{b}(x_t^b, \hat{x}_{t}^b) + \lambda_s \mathcal{L}_\text{sc}(x_t^{sc}, \hat{x}_{t}^{sc})]
\end{equation}

\noindent where $\mathcal{L}_\text{c}$, $\mathcal{L}_\text{b}$, and $\mathcal{L}_\text{sc}$ represent the losses for continuous, binary, and semi-continuous features, respectively, and $\lambda_c$, $\lambda_b$, and $\lambda_s$ are the corresponding weight factors. We empirically observed that this loss function and separate action embeddings are key for ensuring control input sensitivity and correct predictions of the semi-continuous features, as shown in the counterfactual experiments on live network data in \cref{sec:data}.

\subsection{Planning and Control}

We propose to formulate the problem of mobile network control as a receding horizon control problem \cite{mattingley2011receding} and demonstrate it in the context of saving energy by maximizing sleep-time while maintaining network service requirements. 
The control problem formulation can be expressed as a constraint optimization problem with the goal to minimize a cost function (energy consumption) over a prediction horizon $H$ subject to a constraint (service quality) and subject to satisfying a learned system dynamics:

\begin{align}
\min_{u_t, \ldots, u_{t+H-1}} \, & J = \sum_{i=0}^{H-1} C(x_{t+i},u_{t+i}) \\
\text{s.t. }  & x_{t+i+1} \sim f(\ \cdot\ | X_{t+i}, u_{t+i}), i = 0 \ldots H-1 \\
& \mathbb{P}(g(x_{t+i}, u_{t+i}) > 0) < \delta, \quad i = 1 \ldots H,
\label{eq:mpc}
\end{align}

\noindent where $C$ is the cost function, $x_t$ is the predicted state of the network, $X_t=\{x_{t-T+1},...,x_t\}$ is a sequence of input network states, $u_t$ the control input, $g$ represents a constraint function and $f$ is the learned WM.  Here $\delta$ is a constraint tolerance. The constraint is expressed as a probability to cover the case where the WM is stochastic. We define $g$ such that it is positive (denoting violation) when at least one capacity cell is sleeping and the average throughput in the coverage cell falls below a threshold (\SI{50}{\mega\bit\per\second} or \SI{60}{\mega\bit\per\second}). For our cost function, we use $C(x_t,u_t)=||u_t||_1$ to capture the number of times the capacity cell is awake\footnote{Here $||\cdot||_1$ is the 1-norm, and is calculated for non-negative vectors $x$ as $||x||_1=\sum_{i=1}^kx_i$.}.

Note that the formulation in Equations \eqref{eq:mpc} uses an implicit, uncertainty aware WM to model the function $f$. The optimization problem is thus intractable by exact means. Instead, we employ the cross-entropy method for optimization (CEM)~\cite{Kochenderfer2019,chua2018deep}, in which we iteratively sample potential action sequences by using probabilistic priors on $[u_t,...,u_{t+H-1}]$. At a given iteration, the method produces samples of action sequences according to a probability distribution and simulates their outcome using the world model. 
Each sequence of action yields a trajectory for which we compute a cost equal to $J$ plus large penalties for constraint violations.
For stochastic WMs, we sample multiple trajectories using the TS$\infty$ method from \citeauthor{chua2018deep} which yields multiple trajectories for sequences of actions and captures both epistemic and aleatoric uncertainty from the WM~\cite{chua2018deep}. 
In this case, the cost is averaged over the different trajectories, referred to as particles. Thus, if one of the sampled trajectories has a constraint violation, for example if low throughput values are within the uncertainty bounds, the action sequence yielding that trajectory will get associated to a lower overall cost.
After this step, the algorithm ranks the sequences of actions according to their costs and keeps the top fraction of actions. 
A new probability distribution over action sequences is fitted using the selected top actions, which gives more resampling weight to the sequences with lower cost. At the next iteration, new sequences of actions are sampled from this new distribution and evaluated again.
After a fixed number of iterations, the method returns the best action sequence found so far, and the first action of this sequence, $u_t^*$ is applied to the network. The network state then evolves in time, and the optimization method is repeated. The full pseudocode of the algorithm is provided in the Appendix.

%% file: sections/simulation.tex
In this section, we evaluate the proposed approach in a simulated 5G network, which enables safe, closed-loop testing and diverse data generation for energy-saving scenarios. We describe the simulation setup, assess the world model’s predictive performance, and analyze its closed-loop control results.

\subsection{Simulation Environment}

For the simulation experiments, we used a proprietary network simulator calibrated using live network data. 
The simulated network consists of a hexagonal deployment of \num{7} sites with \num{3} sectors each. Each sector is equipped with multiple frequency bands for LTE (\SI{800}{\mega\hertz}, \SI{1800}{\mega\hertz}, and \SI{2600}{\mega\hertz}), and an NR band (\SI{3.5}{\giga\hertz}) refered to as L08, L18, L26, N35 respectively. The L26 band is the capacity cell being controlled by our algorithm, while the L18 band acts as the coverage cell, triggering cell wake-ups. The high frequency cell wake-ups that would happen in reality are not modeled by the simulator, instead we consider that the cell would stay off and we measure eventual throughput constraint violations.

Given a traffic demand for each user, the simulator estimates a snapshot of the network performance. First, it calculates the signal gains from each cell to each user using propagation models described by \citeauthor{asplund2018}~\cite{asplund2018}. It then estimates PRB utilization as well as user throughput, taking into account features like carrier aggregation across LTE cells. 

To make the traffic demand and the simulation data realistic, we used a dataset from a live 5G network (described in \cref{sec:data}) to calibrate them. In this setting, \num{2000} users are spread uniformly in the map, but the traffic demand distribution is non-uniform across users. Using the work of \citeauthor{lee2014} in ~\cite{lee2014}, we fit a Weibull distribution and generate different traffic hotzones in the simulator. A given total traffic volume is then spread across the hotzones according to their relative weights. This modeling creates a spatial diversity in traffic that is similar to a real network's. Finally, for realistic temporal traffic variations, we sample traffic volume traces from the real network data, and apply a scaling factor such that the distribution of PRB utilization in the simulated data would match the distribution in the real data. 

\subsection{World Model Predictions}
In this section, we describe the training data and the training procedure for our ensemble models. We then evaluate the predictive performance of the proposed approach by comparing regression-specific models with multi-task ones. We further compare deterministic and stochastic models.

\noindent \textbf{Synthetic Training Data:} We generate \num{100} network scenarios corresponding to a random intersite distance between \SI{400}{\meter} and \SI{1000}{\meter}, and a realization of the Weibull distribution. For each network, we sample a \num{9} day traffic volume pattern from the real data at \SI{1}{\hour} granularity (1 simulation step corresponds to \SI{1}{\hour}). At each step, a random on/off action is taken for any of the 21 capacity cells. This dataset contains a variety of energy saving actions as well as realistic traffic variation patterns.

\noindent \textbf{World Model Training:}
The state $x_t$ comprises key performance indicators for the different bands and auxiliary information about the state of the network. The features we use include resource utilization percentage and aggregated user throughput for each of the four bands, the sleep state of the capacity band and a time of day encoding, for a total of 11 features. The control input is a binary variable $u_t \in \{0, 1\}^n$ and represents the sleep action of the capacity cells ($n=1$ in the simulation environment). We normalize our data for training and validation and train the ensembles on bootstrapped datasets to increase their diversity. The basic architecture of the models comprises a transformer backbone for sequence handling, normalization layers and linear layers for embedding states and actions and for the output heads.
The training parameters are reported in the Appendix.

\begin{figure}[htbp]
  \centering
  \includegraphics[width=0.8\linewidth]{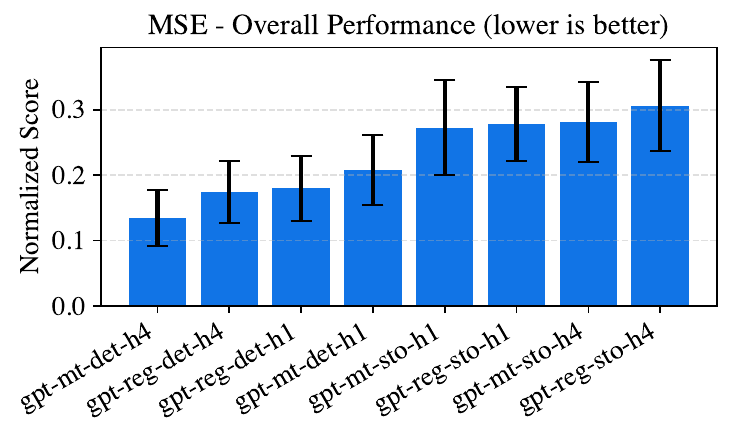}
  \caption{Overall test time prediction errors (mean and standard deviation) averaged over all features and all prediction horizons up to \num{12}.}
  \label{fig:overall_MSE}
\end{figure}

\noindent \textbf{Experiment:}
First, we evaluate the overall performance of the models based on their Mean Squared Error (MSE) as shown in \cref{fig:overall_MSE}. We use MSE for evaluation since, when using the world models for predictions, it is often more important to avoid large errors than to ensure that the mode of the error distribution is small. We compare different model options based on whether they follow the multi-task (MT) architecture or the regression (REG) one, whether they are stochastic or not and finally what horizon they have been trained on. For all the models, we use the weights of the epoch with the best validation score and the episodes we use for testing have not been encountered during training. The model ablations are listed in \cref{tab:ablations}. In this experiment, we calculate the MSE per feature for horizons between \num{1}-\num{12} steps ahead. Then, we average the results over all features and all horizons and report the mean and standard deviation of the error. Since the features have different scales, we normalize the MSE errors for every feature between $[0,1]$ to make them comparable.

\begin{table}[ht]
\centering
\caption{Model Ablations}
\label{tab:ablations}
\begin{tabular}[width=0.9\linewidth]{lcccc}
\toprule
\textbf{Name}  & \textbf{Task} & \textbf{Stochastic} & \textbf{Horizon} \\
\midrule
gpt-mt-det-h4      & MT  & \xmark & 4 \\
gpt-mt-det-h1      & MT  & \xmark & 1 \\
gpt-reg-det-h4     & REG & \xmark & 4 \\
gpt-reg-det-h1     & REG & \xmark & 1 \\
gpt-mt-sto-h4      & MT  & \cmark & 4 \\
gpt-mt-sto-h1      & MT  & \cmark & 1 \\
gpt-reg-sto-h4     & REG & \cmark & 4 \\
gpt-reg-sto-h1     & REG & \cmark & 1 \\
\bottomrule
\end{tabular}
\end{table}

From \cref{fig:overall_MSE}, we see that the multi-task deterministic model with a \num{4}-step horizon (\texttt{gpt-mt-det-h4}) is outperforming the rest of the models. The regression deterministic models do not seem to be improved much by the training horizon increase, while the multi-task deterministic model with a \num{1}-step horizon is worse than its \num{4}-step counterpart. Lastly, we see that the stochastic models have higher validation MSE than their deterministic counterparts. This is not completely unexpected, as stochastic models represent distributions and try to model aleatoric noise to balance fit and uncertainty calibration. This can reduce overconfidence, but it can also introduce noise in the predictive mean. In contrast, deterministic models minimize pointwise errors directly, which yields lower MSE scores. 


\subsection{Closed loop experiments}
In this section, we investigate whether WM-based control is better than a heuristic baseline and a model-free RL algorithm for our network management energy saving use case. We examine the closed loop performance when using CEM with the trained WM and show that we can vary the cost function without a need for retraining. Furthermore, we analyze the effect of using a stochastic versus a deterministic model within the ensembles and whether the uncertainty quantification provides advantages for the closed-loop behavior.\smallskip

\noindent \textbf{Experiment:} We generate \num{10} evaluation scenarios by using the same procedure we used for generating the training data, while ensuring that those episodes are unseen in training. In these evaluation scenarios, we compare the performance of different energy saving control method that can act on the network every hour for a \num{4} day long episode.

\noindent \textbf{Baseline:} The baseline algorithm is a threshold-based method. To turn off the capacity cells, it sets the sleep threshold to \num{0.1} so that when their utilizations fall below the threshold, they will be turned off. Another wake-up threshold of \num{0.25} is used to turn on the capacity cell on again. Correspondingly, wake-up is triggered when the utilization of the coverage cell is above the threshold. The baseline is only active between 1 a.m. and 7 a.m. which is a standard deployment in live networks, thus it can only achieve a maximum of \SI{6}{\hours} per day.

\noindent \textbf{Soft-Actor Critic:} We trained an RL agent using the soft-actor critic (SAC) algorithm \cite{haarnoja2018}. The algorithm was trained for approximately \num{20000} steps (\num{208} scenarios of \num{4} days each) in the same simulation environment used to generate the training data for the WM. For the reward function we used a simple scalarization of \cref{eq:mpc} where the constraint penalty corresponds to a negative reward of CP magnitude in case of constraint violation. Varying CP leads to more or less aggressive energy saving policies which we showcase in our experiments. The policy uses the same input features and action space as the one used by the WM-based controller, as discussed in \cref{sec:approach}.

\begin{figure}
    \centering
    \includegraphics[width=\columnwidth]{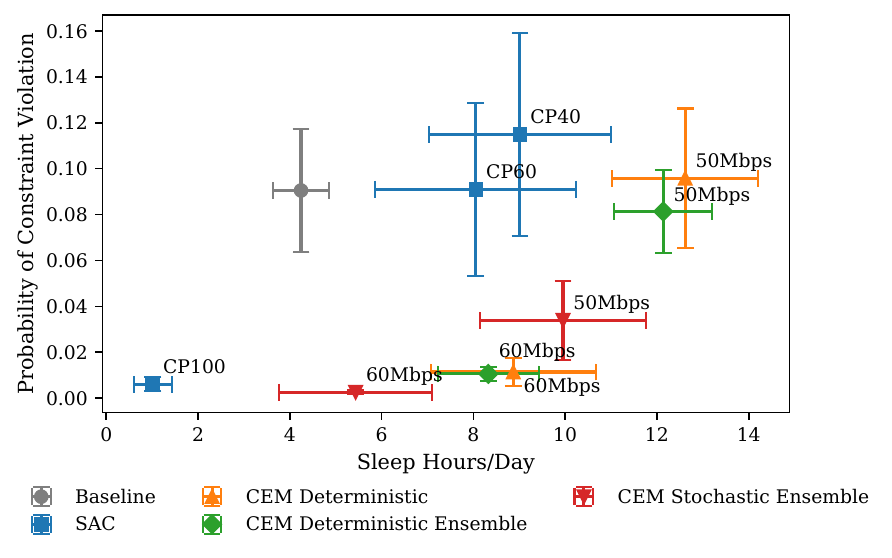}
    \caption{Closed loop performance using \texttt{gpt-mt-sto-h4} and \texttt{gpt-mt-det-h4} for stochastic (total uncertainty) and deterministic (only epistemic) ensembles, respectively. We also test a non-ensemble deterministic baseline, “CEM Deterministic”, without uncertainty quantification. All baselines are examined across different throughput constraints (\SI{50}{\mega\bit\per\second}, \SI{60}{\mega\bit\per\second} for CEM and $\text{CP}=40, 60, 100$) for SAC.}
    \label{fig:sim-perf}
\end{figure}

\noindent \textbf{CEM} CEM is the optimization basis for our proposed WM-based network control approach. We analyze model variants that estimate total uncertainty (stochastic ensemble), only epistemic uncertainty (deterministic ensemble) and no uncertainty (deterministic), and study the impact of changing the level of the constraint in \cref{eq:mpc}. The constraint levels used for the variants are denoted as labels in \cref{fig:sim-perf}, e.g. \SI{50}{\mega\bit\per\second}. Contrary to the RL approach, the change in objective can be directly expressed as a throughput value instead of an abstract penalty and can be changed without retraining. The optimization problem is also scalarized like in the RL agent with $CP=10$ and we do not perform any tuning. We use \num{4} iterations of CEM with \num{100} samples, and keep the top \SI{20}{\percent} of sampled actions across iterations. The execution time is several seconds using a GPU. When using the stochastic WM we sample \num{40} particles and modify the CEM logic to aggregate costs over particles~\cite{chua2018deep}.

We compare the performance of these controllers in terms of average sleep hours per day across episodes and across sectors, as well as the percentage of constraint violations in \cref{fig:sim-perf}. A constraint violation happens when the controller turns off the capacity cells, and this decision is followed by the average user throughput falling below \SI{50}{\mega\bit\per\second} in the L18 coverage cells. We purposely avoid reporting energy saving estimates, as they highly depend on the power consumption capabilities of the different hardware and on operator-specific configurations. Instead, we report the increase in sleep time as a proxy which is easily measurable and could be used with a specific radio hardware power model to estimate actual energy saving.

Note that the impact of an energy saving action on neighboring cells is being modeled by the simulator but is not being observed by any of the methods, thus some violations may be caused by an action happening in neighboring sites. We believe that adding neighbor information could greatly improve the performance of both the RL agent and the WM based control, but this is left as future work.

\begin{figure}
    \centering
    \includegraphics[width=\columnwidth]{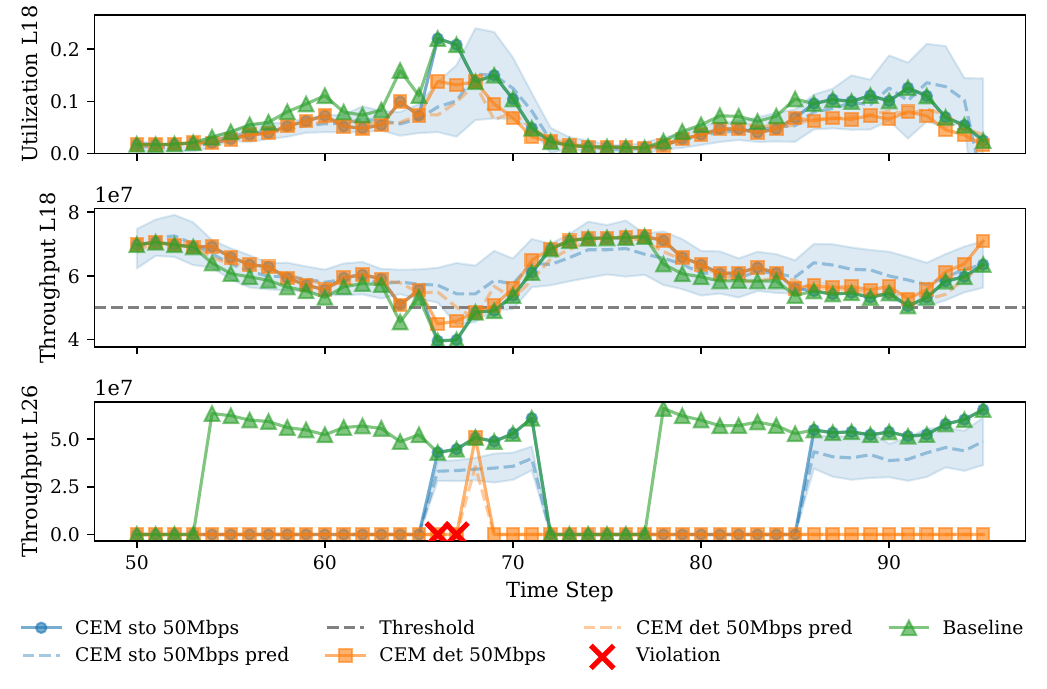}
    \caption{Example of a closed-loop simulation for a sector. We run the same episode using the CEM controller with two different world models, one with uncertainty quantification and one without. The crosses indicate constraint violations. The shaded area is the variance estimated from \num{40} different particles sampled from the stochastic model.}
    \label{fig:cl-trajectory}
\end{figure}

\Cref{fig:sim-perf} shows sleep hours per day each controller achieves, and the probability they will violate the performance constraint. The optimal place to be is the bottom right corner, with max sleep and zero probability of constraint violation. 
We see that the stochastic models are more conservative than the deterministic ones, as they take into the total uncertainty and may detect more potential constraint violations when the algorithm samples multiple trajectories to evaluate each action. Between the deterministic ensembles and single models, we see that estimating epistemic uncertainty only leads to also more conservative sleeping, but with lower violation potential. When tuning the CP parameter for the RL agent's reward, we are able to navigate the trade-off between safety and energy saving. However,  the RL methods are always Pareto dominated by one of the CEM methods. For CEM, we observe the same effect of navigating the trade-off by enforcing higher throughput thresholds than the ones measured in the datasets.

We visualize an example of an evaluation episode for the CEM controllers and the baseline in \cref{fig:cl-trajectory}. We can see that the deterministic model is sleeping more aggressively (L26 throughput of \SI{0}{\mega\bit\per\second}) but fails to predict the L18 throughput violation and only predicts it several steps later, leading to two constraint violations. Thanks to the uncertainty estimate, the stochastic model is able to capture the throughput decrease early on and wake up the capacity cell at the right time. As expected, the baseline is only triggering sleep at narrow time intervals corresponding to night hours. Although the same random seed is used to have the same network scenario, the ground truth values for each method are a little different due to the difference in actions. In general, throughout our experiments we noticed that throughput was one of the most difficult feature to predict, possibly because of the lack of neighbor information and how much it is affected by them.

In conclusion, the presented results show that WM-based control offers a powerful paradigm for optimally changing network configurations and outperforms heuristic baselines and even RL methods. Furthermore, we highlighted the benefit of uncertainty modeling. In the next section, we will assess the performance of our approach on real data.



    



%% file: sections/real_data.tex
In this section, we present experiments based on a proprietary dataset from a live 4G/5G network. We first evaluate the prediction performance of model ablations. 
Then, we qualitatively illustrate the benefit of the multi-task modeling in terms of sensitivity to control actions by showing that regression models tend to ignore the effect of actions. 
Finally, since we cannot directly evaluate the controller actions on a real network, we provide an estimate of the performance based on counterfactual actions. The models are trained as in \cref{sec:simulation}, but for this experiment, we only compare models that have been trained for a prediction horizon~\num{4}. The key hyperparameters are reported in the Appendix.

The dataset consists of 6 months of data from a small live network area with \num{150} sectors. Each sector is equipped with a combination of the following frequency bands
 for LTE: \SI{900}{\mega\hertz}, \SI{1800}{\mega\hertz}, \SI{2100}{\mega\hertz} and \SI{2600}{\mega\hertz}, and two NR bands \SI{700}{\mega\hertz}, \SI{3.5}{\giga\hertz} referred to as L09, L18, L21, L26, N07, N35 respectively. 
 The network state comprises sequences of normalized KPIs for resource utilization, throughput, and traffic volume for the frequency bands, the time of day, the sleep state and the actions of the last 48 hours. For the action, we assume that both bands L21 and L26 can be controlled through on/off actions. In practice, those actions would be converted to CSM thresholds, as explained in~\cref{sec:system}.

\begin{figure}[htbp]
  \centering
  \includegraphics[width=0.85\columnwidth]{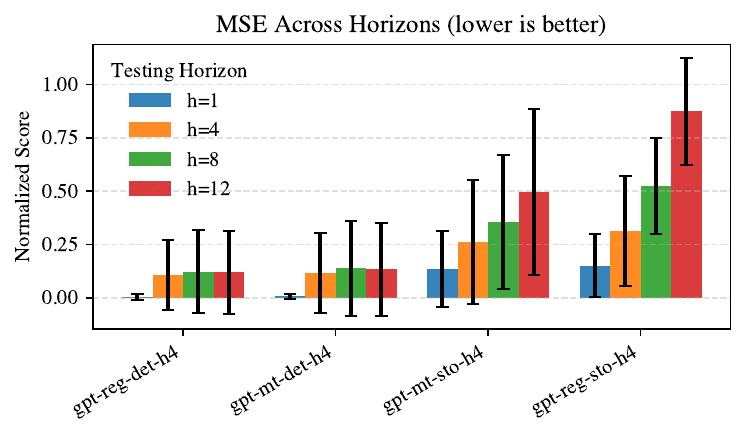}
  \caption{Test time prediction errors (mean and standard deviation) for the real-network data. Errors are averaged over all features.}
  \label{fig:NET_MSE}
\end{figure}

In \cref{fig:NET_MSE} we present the average MSE over all features for different values of the test prediction horizon for the real data. For this dataset, the deterministic multi-task model is not outperforming the regression one, but rather has comparable performance. Consistently with the simulation results, the stochastic models have worse performance. This is to be expected, given that real-network data are significantly noisier than their simulated counterparts. 

\subsection{Counterfactual Actions}

The main advantage of multi-task modeling over standard regression for the deterministic WMs becomes more pronounced when we look at their sensitivity to the control input. 
We compute counterfactual actions using the CEM algorithm on data corresponding to one week, by replaying the network historical data. 
The counterfactual actions correspond to energy saving actions that would have been applied to the network if the CEM algorithm were deployed closed loop.

\begin{figure}
    \centering
    \includegraphics[width=\columnwidth]{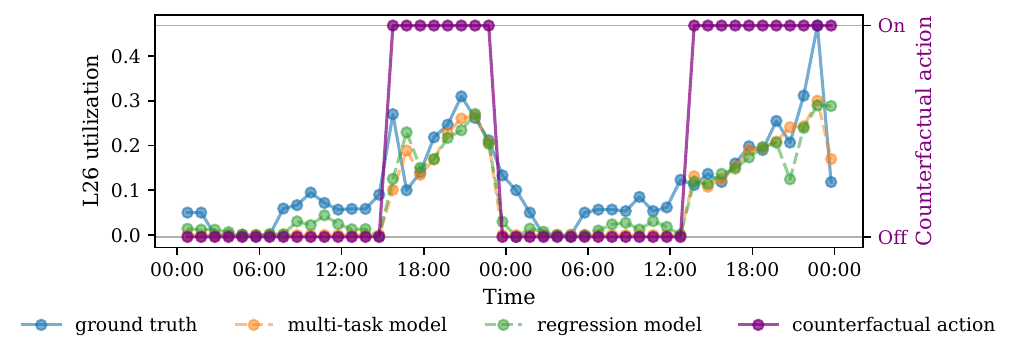}
    \caption{Example of one sector's LTE capacity utilization evolution. In purple, we visualize a counterfactual action proposed by our WM-based controller. The blue line corresponds to the ground truth utilization, while the orange and green lines correspond to WM predictions for the counterfactual action.}
    \label{fig:counterfactual}
\end{figure}

In \cref{fig:counterfactual}, we show counterfactual actions, that the controller would have applied to the network, on the right axis. We compare the predictions of the multi-task WM and the ones of the regression WM, when using the counterfactual action as input instead of the ground truth actions from the dataset without changing any of the other input KPIs. The multi-task WM is able to correctly predict that given the “off” action, the utilization of the capacity carrier would go to zero. However, the regression WM is predicting a value close to the ground truth and shows little sensitivity to the control action, making it a poor candidate for WM-based control.

\subsection{Open-loop Evaluation}

We extend the analysis of the counterfactual actions by estimating the open-loop performance of the controller. We first estimate the number of extra sleep hours per day and per carrier averaged over all sectors, as if the controller were deployed. 
We do so by assuming that all “off” actions recommended by CEM on the L21 and L26 cells would have caused them to sleep for an hour. 

Since the actions are not applied, it is not possible to estimate their consequences on user throughput. Instead, we consider the worst case estimate in utilization increase for L18. 
When a capacity cell is set to sleep, we read its current utilization in the data and assume that the whole utilization would be offloaded to the L18 cell of the same sector. 
This PRB transfer is reasonable as the bandwidth is generally the same for the mid-band LTE carriers. In reality, users may also be offloaded to neighboring sectors, or other available coverage bands in the sector.
We will consider as constraint violation every case where the total utilization in the coverage cell would exceed 60\% which in practice would trigger a wake-up by the CSM software. 
A baseline algorithm is running on all sectors of the dataset, and thus we are able to directly compare the effect of CEM on the sleep time as well as its effect on increased utilization.

In \cref{tab:openloop-perf}, we compare the performance of variants of our WM-based controller (using \texttt{gpt-mt-sto-h4} for the stochastic and \texttt{gpt-reg-det-h4} for the deterministic WMs) as we experiment with different throughput constraints (\SI{50}{\mega\bit\per\second} and \SI{60}{\mega\bit\per\second}). Consistent with the simulation results, our WM control method proposes a significant number of sleep hours that decreases with stricter throughput constraints, while the stochastic WM provides more conservative control strategies with less sleep. The large standard deviation in the estimated sleep hours is due to the fact that the dataset contains both sectors with very high average utilization (little sleep potential) and sectors with low utilization (high sleep potential).
The fraction of samples with utilization over 60\% would significantly increase with the CEM methods compared to the baseline, which is potentially risky. This number further increases with the stochastic WMs, which could be a reflection of their poor accuracy or an indication that congestion should be considered in the cost function. For further evaluation, the full empirical distribution of the utilization and counterfactual utilization under different WM controllers can be found in the Appendix. 
Lastly, we see that the CEM controllers lead to more “on/off” switches per day, as they lead to multiple sleeping windows. In cases where the overhead of switching sleep states is significant, a longer control horizon and a penalty for switching would be advisable. 

\begin{table}
\caption{Performance comparison of baseline and CEM algorithms when running open loop on 1 week of the real dataset.}
\label{tab:openloop-perf}
\begin{tabularx}{\columnwidth}{lXXX}
\toprule
\text{\textbf{Algorithm}} & \text{\textbf{Sleep Hours}} \text{(h/day)} & \text{\textbf{Fraction with}} \text{\textbf{Util.$\geq$60\%}} & \text{\textbf{Switches}} \text{(per day)} \\
\midrule
Baseline & 4.5 $\pm$ 0.9 & 0.004 $\pm$ 0.019 & 1.9 $\pm$ 0.3 \\
CEM-det-50M & 18.4 $\pm$ 6.4 & 0.019 $\pm$ 0.049 & 3.3 $\pm$ 2.1 \\
CEM-det-60M & 13.4 $\pm$ 7.8 & 0.011 $\pm$ 0.038 & 3.7 $\pm$ 2.2 \\
CEM-sto-50M & 16.0 $\pm$ 6.9 & 0.022 $\pm$ 0.050 & 4.1 $\pm$ 2.4 \\
CEM-sto-60M & 11.0 $\pm$ 7.4 & 0.015 $\pm$ 0.039 & 4.3 $\pm$ 1.8 \\
\bottomrule
\end{tabularx}
\end{table}

\begin{figure}
    \centering
    \includegraphics[width=\columnwidth]{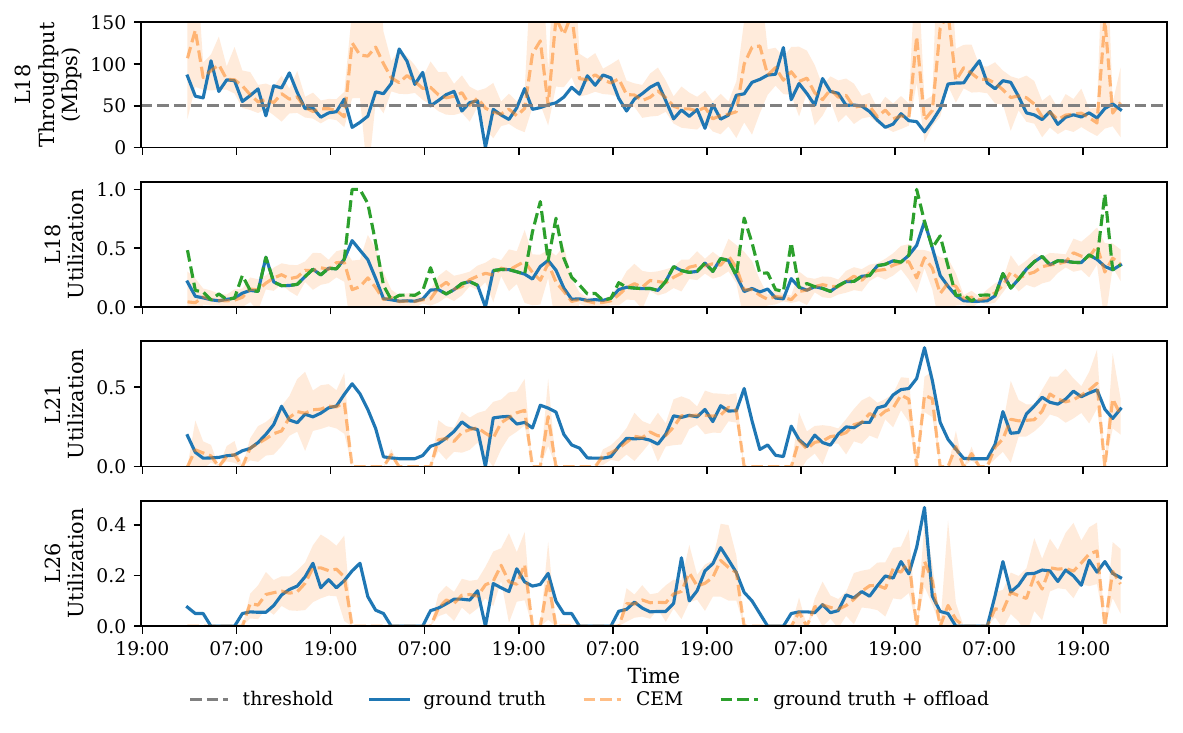}
    \caption{Example of running the CEM policy with a stochastic model on a sample of the real data. In orange, we visualize the prediction by using the counterfactual actions as input. The shaded area is the variance estimated from \num{40} different particles sampled from the stochastic model. The green dashed line represent the offloaded utilization assuming a worst case scenario.}
    \label{fig:openloop}
\end{figure}

Finally, we visualize actions and predictions from the WM-based controller in \cref{fig:openloop}. The multi-task model is able to predict zero utilization when an “off” action is suggested, while predicting the value accurately with an “on” action, demonstrating its ability in handling semi-continuous variables.
Note that the controller is acting on two frequency bands here, and it proposes a significant increase in sleep time (samples with 0 utilization) compared to the historical data.
In the second row, we visualize the hypothetical offloaded utilization on the coverage layer with a dashed line and  see that a few events of high utilization would occur.

In this analysis, we demonstrated the potential of WM-based control by training a model and evaluating it on real data. Contrary to simulation data, it is more challenging to train stochastic WMs and WMs that are sensitive to actions. With our proposed multi-task modeling, the controller proposes reasonable actions. Our estimation of the offloaded utilization suggests that further improvement of the model predictions and controller constraints may be needed before a field deployment. However, with reasonable safety check in place, such as implementing “off” actions with CSM thresholds, the use of this method may bring significant increase in sleep times.

%% file: sections/conclusion.tex
The increasing usage and complexity in mobile metworks, is making their intelligent and automated control crucial. In this paper, we propose a world model-based approach for network control that decouples model learning from the controller itself, allowing the same model to be reused for different objective functions. The proposed method applies to various input feature types, estimates its own uncertainty, and eliminates the need for retraining when the optimization objective changes.
With CEM, we have demonstrated how this approach can control network parameters in a receding-horizon fashion, exemplified by cell sleep mode control. Our experiments showed that the proposed framework Pareto-dominates both rule-based baselines and state-of-the-art reinforcement learning benchmarks in closed-loop simulations, while also showing strong potential in open loop emulation of live network data.
Future work will explore the integration of network graph topology into the prediction models to provide richer neighbor information, the use of network foundation models to improve generalization and accuracy, and the evaluation of closed-loop deployments of the world model-based control in live network scenarios.

%% file: sections/appendix.tex
\appendix
In this appendix, we provide further training details on the world model for reproducibility and an extra evaluation of the model ablations. Furthermore, we describe how CEM works in pseudocode and present the empirical cumulative distribution function of utilization for the real-world, open-loop evaluation.

\subsection{World-model Details}

\noindent{\textbf{Auto-regressive training}}
All the models that have been used in this work are sequence to sequence models. Namely,
given input sequence $X_t=\{x_{t-T+1}, ..., x_t\}$ and action sequence $U_t =\{u_{t-T+1}, ..., u_{t+T+H-1}\}$, where $T$ is the input history length and $H$ is the prediction horizon, our models predict $\hat{Y}_t = \{\hat{x}_{t-T+2}, ..., \hat{x}_{t+H}\}$, with auto-regressive predictions for steps $t+2$ to $t+H$. To improve training stability, all the models are trained with teacher forcing \cite{teacher-forcing}. 

\noindent{\textbf{Model architecture}}
The basic architecture of the models comprises the following layers: a RevIn layer~\cite{Kim2022ReversibleIN}, linear layers for embedding the states and actions separately, a layer norm layer~\cite{ba2016layernormalization}, a GPT-2 model for handling the sequences of data and linear output heads. A RevIn layer normalizes the data within their own part of the timeseries, addressing potential distribution shift problems where different segments may have varying scales and statistical properties. 
We used separate linear layers for embedding the states and actions as a way of learning feature representations that preserve the importance of the control input. These are followed by a layer normalization layer, which stabilize training when working with transformer-based architectures that are sensitive to gradient flow.
We chose a GPT-2 transformer as the sequence handling backbone, due to its proven ability to capture long-range dependencies and complex temporal patterns through its self-attention mechanism. We also experimented with an LSTM backbone, but the transformer model seemed to outperform it. 

\begin{figure}
    \centering
    \includegraphics[width=0.85\columnwidth]{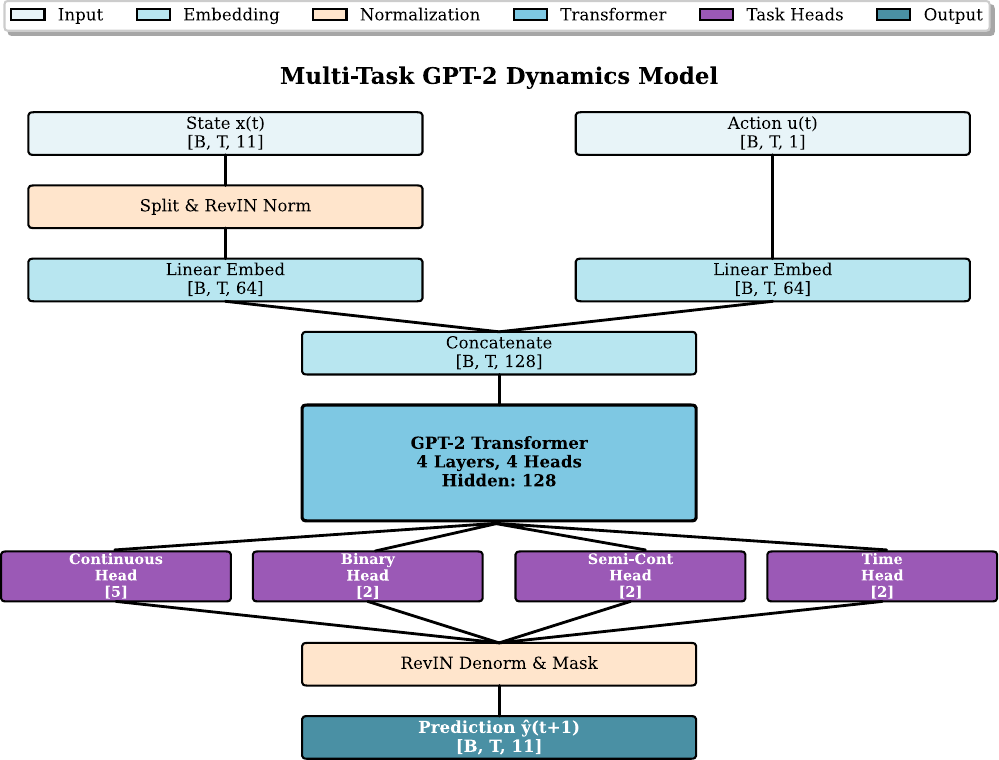}
    \caption{The architecture for the \texttt{gpt-mt-det} model.}
    \label{fig:gpt-mt_architecture}
\end{figure}

Linear output heads provide the final mapping from the transformer's learned representations to the target prediction space. The specific structure of the linear heads depends on whether the model is stochastic or deterministic, and whether it is used for regression or as a multi-task model. The stochastic models have twice the output neurons as they output the mean prediction and its uncertainty, while the multi-task models have specific “task” heads for every feature category. Finally, the predictions are denormalized through the RevIn layer, so the outputs are in the original data scale, while maintaining the benefits gained from normalized training.
\cref{fig:gpt-mt_architecture} illustrates the above architecture for the case of the deterministic multi-task model.

\noindent{\textbf{Uncertainty propagation}}
In \cref{sec:approach}, we described the basics of uncertainty modeling with stochastic and deterministic ensembles. Here, we provide more details about how the total uncertainty is modelled in each case, starting from the stochastic regression models. In this case, we assume that the predictive distribution of the (continuous) variables is a Gaussian, so the output of the ensemble is also a Gaussian whose mean and variance is the result of the equally weighted predictions from the $M$ ensemble members:
\begin{align}
\bar{\mu} &= \frac{1}{M} \sum_{i=1}^{M} \mu_i \\[6pt]
\sigma_{\text{ensemble}}^2 &= 
\underbrace{\frac{1}{M} \sum_{i=1}^{M} \sigma_i^2}_{\text{Aleatoric uncertainty}}
\;+\;
\underbrace{\frac{1}{M} \sum_{i=1}^{M} (\mu_i - \bar{\mu})^2.}_{\text{Epistemic uncertainty}}
\label{eq:total_uncertainty}
\end{align}

In practice, to propagate the uncertainty during the predictions, for every step we create a Gaussian distribution around the models' predicted mean and variance and sample from it using the reparametrization trick. The sampled prediction is then used as the auto-regressive input for the next step. Note that for numerical stability during training, we actually predict the \emph{log variance} (clipped to the range $[-10, 10]$) and then transform it as needed to variance or standard deviation. 

In the case of deterministic regression models, the predictions are single points instead of Gaussians and their variances are zero, so the ensemble estimates only the epistemic part of \cref{eq:total_uncertainty}. During the auto-regressive predictions, we do not sample any points, we just propagate the point estimate of the mean.

Special care should be taken for the case of uncertainty propagation with the stochastic multi-task models, as only some features are continuous and can be modelled by Gaussians the same way as described above. For the binary features, we instead assume a Bernoulli distribution and model its variance as $\sigma = p(1-p)$, where $p$ is the predicted probability that the binary task head outputs. In the instantiations of the models we compare in the experiment section, we do not further sample around the binary prediction. Lastly, the semi-continuous features are treated like the continuous ones, but the sampled predictions are masked according to the binary predictions.

\begin{table}[ht]
\centering
\caption{Training parameters for the world model experiments.}
\label{tab:training_params}
\begin{tabularx}{\columnwidth}{lXX}
\toprule
\textbf{Parameter} & \textbf{Simulated data} & \textbf{Real data} \\
\midrule
State input length & 48 timesteps & 48 timesteps \\ 
Train–test split & 90–10\% & 95-5\%   \\
Number of training episodes & 900 days  & 157 days \\
Training prediction horizon & 1 or 4 steps  & 4 steps\\ 
Test prediction horizon & 12 steps & 12 steps \\
Number of ensemble members & 5  & 5\\
Optimizer & Adam  & Adam \\
Learning rate & $1\times10^{-4}$ & $1\times10^{-4}$ \\
Epochs & 30 & 50 \\
Training time (GPU) & 2–8 hours & 6–16 hours \\
\bottomrule
\end{tabularx}
\end{table}

\begin{figure}
  \centering
  \includegraphics[width=0.95\linewidth]{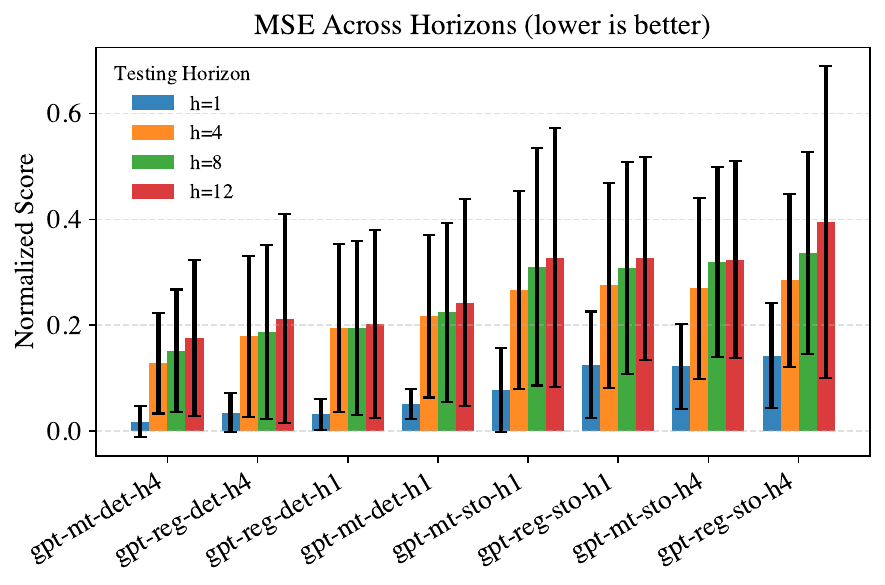}
  \caption{Test time prediction errors (mean and standard deviation) averaged over all features for different test prediction horizons.}
  \label{fig:group_MSE}
\end{figure}

\noindent{\textbf{Prediction Errors Per Horizon:}}
In \cref{fig:group_MSE} we visualize model prediction errors for different test  horizons, averaged over all features. The performance of most models is not very consistent across features, which leads to large error- bars over the mean MSE. It is still possible to draw high-level conclusions about how the MSE scales with the test prediction horizon, however to choose a model for specific applications it is recommended to study individual features according to their importance.
Nevertheless, the results per test prediction horizon are consistent with that of \cref{fig:overall_MSE} but we additionally see that for the deterministic models, extending the training prediction horizon to \num{4}, improves performance for training horizon \num{1} as well. For the multi-task stochastic models (\texttt{gpt-mt-sto-h1, gpt-mt-sto-h4}), we see that the horizon \num{4} model has worse performance for horizon \num{1} but slightly better, and more consistent, behavior for the other horizons underlying the usefulness of training with a longer horizon. It should be noted that extending the training prediction horizon affects the training time as well. On average, we observed that increasing the horizon from \num{1} to \num{4}, almost doubles the training time.

\subsection{Planning with the Cross Entropy Method}

Here we provide a pseudo-code of the cross entropy method we use to solve \cref{eq:mpc}. We sample different trajectories per action sequences when the WM is a stochastic ensemble. 

\begin{algorithm}[H]
\caption{CEM Planning with Discrete Actions and Ensemble WM}
\label{alg:cem_simple}
\begin{algorithmic}[1]
\STATE \textbf{Input:} state $x_t$, horizon $H$, discrete actions $\mathcal{U}$, cost $C$, constraint $g$, ensemble $\{f_m\}_{m=1}^M$
\STATE \textbf{Hyperparams:} samples $N$, elites $N_e$, iterations $L$, particles $P$, penalty $CP$
\STATE Initialize distribution over action sequences
\FOR{$\ell = 1$ to $L$}
    \FOR{$n = 1$ to $N$}
        \STATE Sample action sequence $\tilde{\bm{u}}^{(n)}$ from current distribution
        \STATE $S^{(n)} \leftarrow 0$
        \FOR{$p = 1$ to $P$}
            \STATE Sample model $m_p$ from ensemble
            \STATE Roll out using $\tilde{\bm{u}}^{(n)}$ and $f_{m_p}$
            \STATE $S^{(n)} \mathrel{+}= \frac{1}{P}\sum_{i=0}^{H-1}\big[C(x_{t+i},\tilde{u}^{(n)}_{t+i})+CP\,\mathbf{1}\{g(x_{t+i},\tilde{u}^{(n)}_{t+i})>0\}\big]$
        \ENDFOR
    \ENDFOR
    \STATE Select $N_e$ sequences with lowest $S^{(n)}$
    \STATE Update distribution parameters using elite sequences
\ENDFOR
\STATE \textbf{Return:} first action of best sequence
\end{algorithmic}
\end{algorithm}

\subsection{Open-loop Evaluation}

\begin{figure}
    \centering
    \includegraphics[width=\columnwidth]{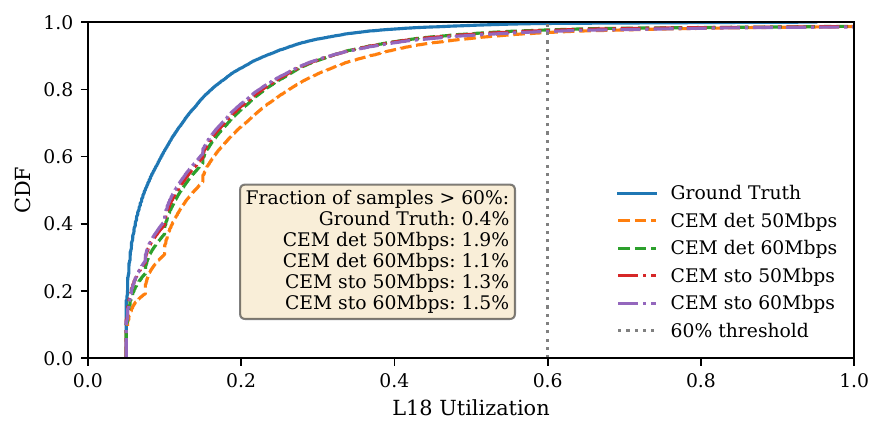}
    \caption{Estimated increase in utilization if the counterfactual actions proposed by CEM were applied in the real system, assuming the worst case scenario where all traffic is offloaded to the L18 layer.}
    \label{fig:cdf-counterfactual}
\end{figure}

In \cref{fig:cdf-counterfactual}, we show the empirical cumulative distribution function of the utilization over all the samples in the 1 week of data used for the open loop analysis with real data. We can see that the CEM method would cause a big shift of utilization but that the majority of sample would still be below the 60\% and thus hints that a majority of the sleep actions could be valid.
